\documentclass[preprint, 3p, 12pt]{elsarticle}

\usepackage{amssymb}

\usepackage{float}
\usepackage{amsmath}
\usepackage{subfig}
\usepackage{hyperref}
\usepackage[export]{adjustbox}

\journal{Acta Biomaterialia}

\begin{document}

\begin{frontmatter}



\title{A Gaussian process approach for rapid evaluation of skin tension}


\author[inst1,inst2]{Matt Nagle\corref{cor1}}
\ead{matt.nagle@ucdconnect.ie}

\author[inst2]{Hannah Conroy Broderick}

\author[inst2,inst3]{Christelle Vedel}

\author[inst2,inst4]{Michel Destrade}

\author[inst5]{Michael Fop\fnref{flag1}\corref{cor1}}
\ead{michael.fop@ucd.ie}

\author[inst2,inst6]{Aisling Ní Annaidh\fnref{flag1}\corref{cor1}}
\ead{aisling.niannaidh@ucd.ie}

\fntext[flag1]{These authors contributed equally to this work}
\cortext[cor1]{Corresponding authors at: School of Mechanical and Materials Engineering, UCD, Belfield, Dublin 4, Ireland. School of Mathematical and Statistical Sciences, UCD, Belfield, Dublin 4, Ireland}

\affiliation[inst1]{organization={SFI Centre for Research Training in Foundations of Data Science, University College Dublin},
            addressline={Belfield}, 
            city={Dublin 4},
            country={Ireland}}

\affiliation[inst2]{organization={School of Mechanical and Materials Engineering, University College Dublin},
            addressline={Belfield}, 
            city={Dublin 4},
            country={Ireland}}

\affiliation[inst3]{organization={EPF School of Engineering},
            addressline={Av. du Président Wilson}, 
            city={Cachan},
            country={France}}

\affiliation[inst4]{organization={School of Mathematical and Statistical Sciences, University of Galway},
            addressline={University Rd},
            city={Galway},
            country={Ireland}}

\affiliation[inst5]{organization={School of Mathematics and Statistics, University College Dublin},
            addressline={Belfield}, 
            city={Dublin 4},
            country={Ireland}}

\affiliation[inst6]{organization={Charles Institute of Dermatology, University College Dublin},
            addressline={Belfield}, 
            city={Dublin 4},
            country={Ireland}}

\begin{abstract}
Skin tension plays a pivotal role in clinical settings, it affects scarring, wound healing and skin necrosis. Despite its importance, there is no widely accepted method for assessing \textit{in vivo} skin tension or its natural pre-stretch. This study aims to utilise modern machine learning (ML) methods to develop a model that uses non-invasive measurements of surface wave speed to predict clinically useful skin properties such as stress and natural pre-stretch. A large dataset consisting of simulated wave propagation experiments was created using a simplified two-dimensional finite element (FE) model. Using this dataset, a sensitivity analysis was performed, highlighting the effect of the material parameters and material model on the Rayleigh and supersonic shear wave speeds. Then, a Gaussian process regression model was trained to solve the ill-posed inverse problem of predicting stress and pre-stretch of skin using measurements of surface wave speed. This model had good predictive performance ($R^2$ = 0.9570) and it was possible to interpolate simplified parametric equations to calculate the stress and pre-stretch. To demonstrate that wave speed measurements could be obtained cheaply and easily, a simple experiment was devised to obtain wave speed measurements from synthetic skin at different values of pre-stretch. These experimental wave speeds agree well with the FE simulations and a model trained solely on the FE data provided accurate predictions of synthetic skin stiffness. Both the simulated and experimental results provide further evidence that elastic wave measurements coupled with ML models are a viable non-invasive method to determine \textit{in vivo} skin tension.\\

\noindent
\textbf{Statement of Significance:}\\
To prevent unfavorable patient outcomes from reconstructive surgery, it is necessary to determine relevant subject-specific skin properties. For example, during a skin graft, it is necessary to estimate the pre-stretch of the skin to account for shrinkage upon excision. Existing methods are invasive or rely on the experience of the clinician. Our work aims to present a novel framework to non-invasively determine \textit{in vivo} material properties using the speed of a surface wave traveling through the skin. Our findings have implications for the planning of surgical procedures and provides further motivation for the use of elastic wave measurements to determine \textit{in vivo} material properties.
\end{abstract}

\begin{graphicalabstract}
\includegraphics[width=\textwidth]{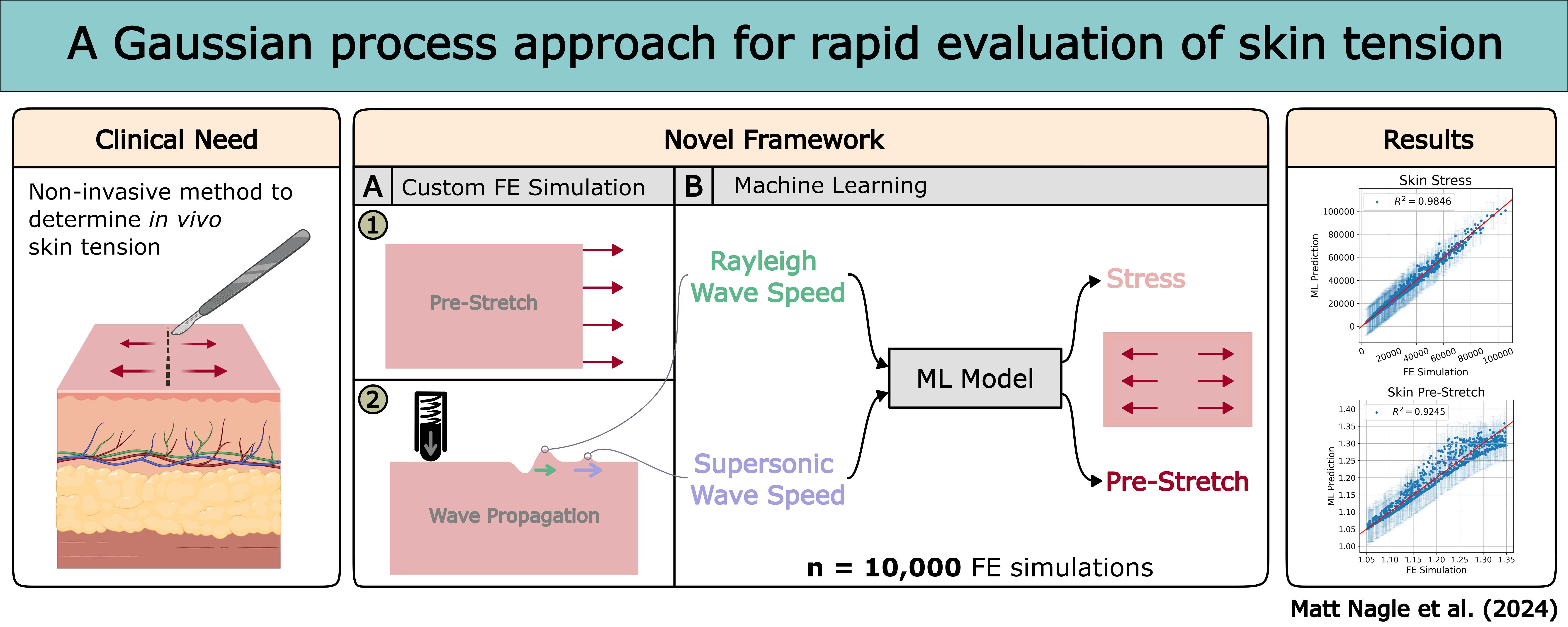}
\end{graphicalabstract}



\end{frontmatter}



\section{Introduction}\label{Introduction}

The skin is the largest organ in the body and serves as the interface between the internal physiological environment and the external world. It plays a pivotal role in protection against external threats including the invasion of pathogens and fending off chemical and physical assaults \cite{proksch_skin_2008}. It is under constant anisotropic tension and must be both pliable and durable for everyday movement.

In the surgical setting, skin tension plays a crucial role in achieving optimal outcomes and fostering effective wound healing processes \cite{harn_tension_2019} \cite{paul_biodynamic_2018}. In many surgical procedures, such as wound closure or breast reconstructions, maintaining appropriate skin tension is paramount. Excessive tension can lead to complications such as compromised aesthetics and scarring \cite{paul_biodynamic_2018} \cite{son_overview_2014}, and further, to significant psychosocial impacts for the patient \cite{natalia_ziolkowski_psychosocial_2019}. It can also lead to dangerous physical complications including wound dehiscence, hematoma and skin necrosis, which can occur at alarmingly high complication rates for patients. For example, recent publications have reported overall complication rates for head and neck tissue expansion of 8.73\% \cite{azzi_complications_2020} and that 8.9\% of patients experience skin necrosis from breast tissue expansion following a mastectomy \cite{yalanis_mastectomy_2015}.

It is well established that \textit{in vivo} skin tension is aligned along preferred directions known as Langer lines, skin tension lines or relaxed skin tension lines \cite{langer_anatomy_1978} \cite{borges_relaxed_1984} \cite{paul_biodynamic_2018}. Both the magnitude of \textit{in vivo} skin tension and its preferred orientation have been shown to be patient-specific \cite{deroy_non-invasive_2017} \cite{nagle_analysis_2023}. However, despite the important role skin tension plays in surgery, there is no commonly accepted quantitative method to determine its magnitude or direction \textit{in vivo}. Currently, surgeons must rely on generic skin tension maps or an imprecise ``pinch test'' to identify the orientation of skin tension lines, which requires significant skill and experience to interpret \cite{deroy_non-invasive_2017} \cite{ni_annaidh_tension_2019} \cite{seo_estimating_2013}.

Recently, attempts have been made to identify the \emph{direction} of skin tension lines using suction devices \cite{laiacona_non-invasive_2019}, extensiometry \cite{boyer_dynamic_2009} \cite{paul_biodynamic_2017} and elastic wave propagation \cite{deroy_non-invasive_2017} \cite{ruvolo_skin_2007}. There have also been attempts to quantify the \emph{magnitude} of \textit{in vivo} pre-stretch and skin tension. However, many of the methods are invasive and cumbersome and have not been widely adopted \cite{dauendorffer_shrinkage_2009} \cite{deroy_non-invasive_2017} \cite{jor_estimating_2011} \cite{paul_new_2016}. 

Most methods to determine the magnitude of pre-stretch involve a destructive process where the skin is excised and the shrinkage is quantified \cite{dauendorffer_shrinkage_2009} \cite{deroy_non-invasive_2017} \cite{jor_estimating_2011}. A notable exception is the method employed by Paul et al. where instead of skin being excised, rods are used to compress or stretch the skin \cite{paul_new_2016}. However, the measurement process is still invasive as the rods must pierce the skin for a measurement to be taken. 

Our own recent publication analysed the direction and relative magnitude of skin tension using a wave propagation device (Reviscometer\textsuperscript{\tiny\textregistered} Model RVM 600, made by Courage \& Khazaka Electronic GmbH) to take \textit{in vivo} measurements of the surface wave speed \cite{nagle_analysis_2023}. Devices such as these can be made easily and cheaply to facilitate measurements of wave speed on the surface of the skin along one axis. They often contain two piezoelectric transducers spaced a known distance apart. One transducer impacts the surface of the material, generating a surface wave, while the other transducer detects the resulting wave and records the time taken for that wave to propagate across the surface of the skin, along one axis. We demonstrated that the direction of highest skin tension and its magnitude is subject specific and is affected by the age and sex of the patient, and that skin tension is directly related to the speed of the elastic wave \cite{nagle_analysis_2023}. We concluded that \textit{in vivo} elastic wave measurements are a suitable method for inferring \textit{in vivo} skin tension. 

There exist analytical models relating the material properties to a surface wave speed. For example, for a Rayleigh surface wave traveling over a Mooney-Rivlin half space under uniform uniaxial tension, the wave speed is \cite{flavin_surface_1963}:
\begin{equation} \label{Modified Flavin Equation}
    v = \sqrt{\frac{E}{6 \rho} \left[ (1 - \beta) {\lambda_1}^{2} + (1 + \beta) \lambda_1 \right] \left( 1 - (0.2956)^2 {\lambda_1}^{-3} \right)},
\end{equation}

where $v$ is the wave speed along one axis, $E$ is the Young's modulus, $\rho$ is the density, $\beta$ is a dimensionless material parameter which ranges from the neo-Hookean case ($\beta = -1$) to the extreme Mooney-Rivlin case ($\beta = +1$), see Section \ref{Input Space Sampling Section} for more details; $\lambda_1$ is the pre-stretch in the direction of tension.

This analytical solution can be a useful tool, but, its real-world use is limited as it makes a number of assumptions and outputs a single wave which travels at a constant velocity. More recently, Li et al. developed an analytical solution that describes two propagating waves to account for the Rayleigh and supersonic shear waves \cite{li_supershear_2022}. Such analytical solutions can be useful for quantifying material parameters of interest using non-destructive means. Notably, Feng et al. developed a traveling-wave optical coherence elastography technique to measure the elastic modulus of the epidermis, dermis, and hypodermis \cite{feng_vivo_2022}. However, to the best of our knowledge, there is no analytical method to determine the magnitude of skin tension, stress or pre-stretch.

It seems that, an objective method has not yet been developed that can non-invasively determine important subject specific parameters such as skin tension. As such, the overall objective of the paper is to present a method which can non-invasively identify the magnitude of \textit{in vivo} skin tension and stress using surface wave speeds. To this goal, we have devised the following procedure, which constitutes the novel contribution of this research:

\begin{enumerate}
    \item Development of a simplified finite element (FE) model that simulates a typical surface wave propagation experiment in \textit{in vivo} skin.
    \item Creation of a large database of simulated test cases representative of real-world conditions.
    \item Development of a statistical emulator for the purpose of a sensitivity analysis to elucidate the general trends and important features of surface wave propagation in \textit{in vivo} skin.
    \item Development of a machine learning (ML) model which can solve, in real time, the ill-posed and inverse problem of determining \textit{in vivo} tension and stress from elastic wave speeds.
\end{enumerate}

\section{Materials and Methods}

\subsection{Finite Element Modelling} \label{FE Section}

The FE model for this study was designed to be both computationally inexpensive, as our analysis will involve running many simulations, and analogous to existing wave propagation devices (for example, the Reviscometer\textsuperscript{\tiny\textregistered}).

\begin{figure}[ht]
  \centering
  \includegraphics[width=\textwidth]{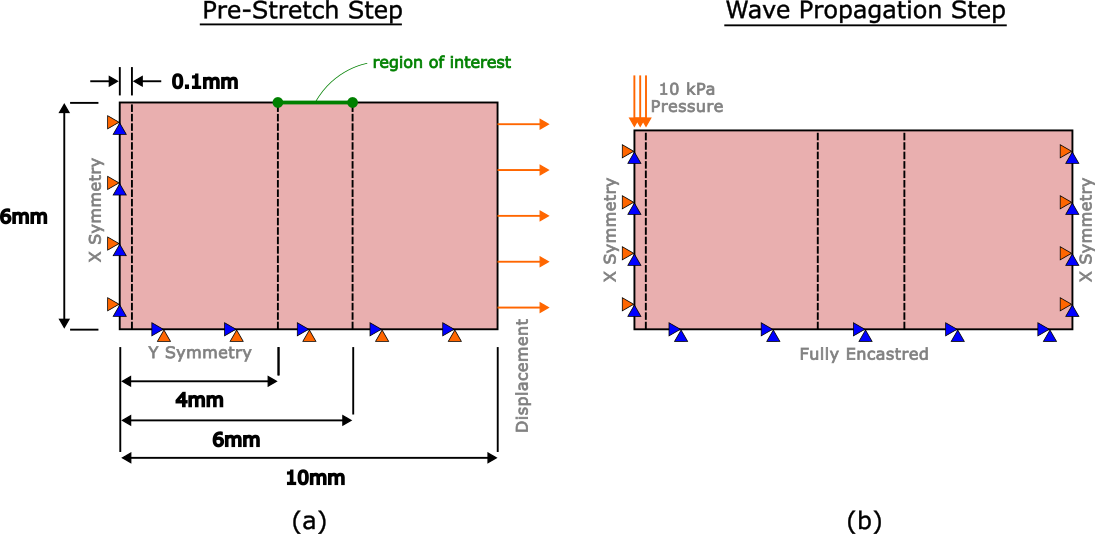}%
  \caption{Dimensions and boundary conditions of the FE model of wave propagation. (a) The uniaxial pre-stretch is generated using a displacement boundary condition and (b) the wave is generated by a 10 kPa pressure applied for $2 \times 10^{-5}$ s. The vertical displacement of the nodes in the $4 \, \text{mm} - 6 \, \text{mm}$ region was stored for analysis.}
  \label{Schematic of FE Model}
\end{figure}

We simulate a pre-stretched two-dimensional block of skin and impact the surface, see Figure \ref{Schematic of FE Model}. This generates a wave that propagates along the surface of the skin. The vertical displacement of the nodes at known distances from the impact site can then be stored for analysis. To implement the model, firstly, the nonlinear FE package Abaqus/Standard (Dassault Systems, Waltham, MA) was used to statically pre-stretch the skin and, subsequently, Abaqus/Explicit (Dassault Systems, Waltham, MA) was used to perform the wave propagation. The unstretched skin block has dimensions $10 \, \text{mm} \times 6 \, \text{mm}$ and was discretised into 150,000 CPS4R elements with 150,801 nodes. The dimensions of the skin block were selected to minimise wave reflections interfering with waveforms from the region of interest (surface nodes $4\,\text{mm} - 6\,\text{mm}$ from the impact). The skin block is modelled by a hyperelastic material (either neo-Hookean or Mooney-Rivlin). 

To perform the pre-stretch, a displacement boundary condition was used to perform a uniaxial stretch. After the pre-stretch, a wave was generated by applying a 10 kPa pressure for $2 \times 10^{-5}$ s; see Figure \ref{Schematic of FE Model}.

\begin{figure}[ht]
    \centering
    \subfloat[]{\label{Typical Waveform Shape:a}\includegraphics[width=.5\linewidth]{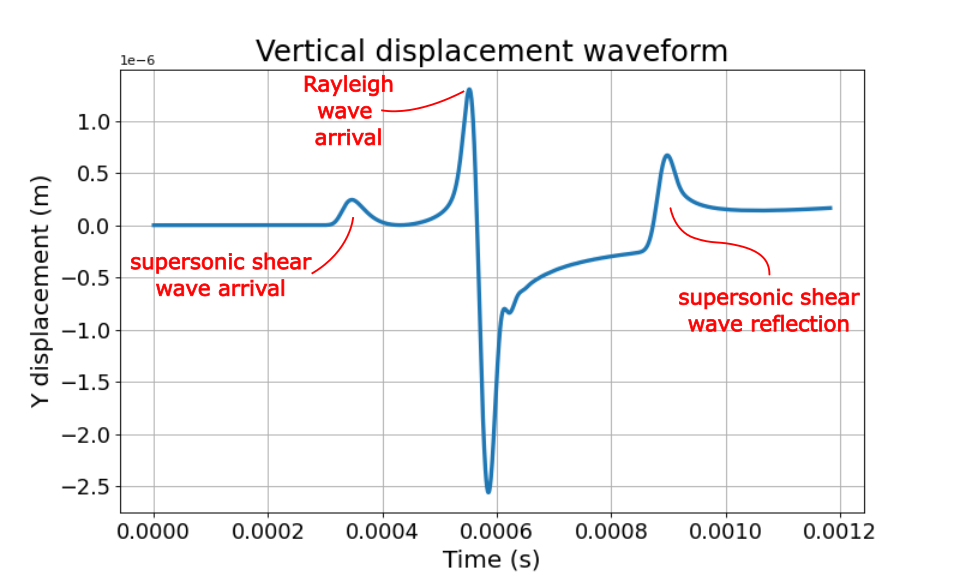}}
    \subfloat[]{\label{Typical Waveform Shape:b}\includegraphics[width=.5\linewidth]{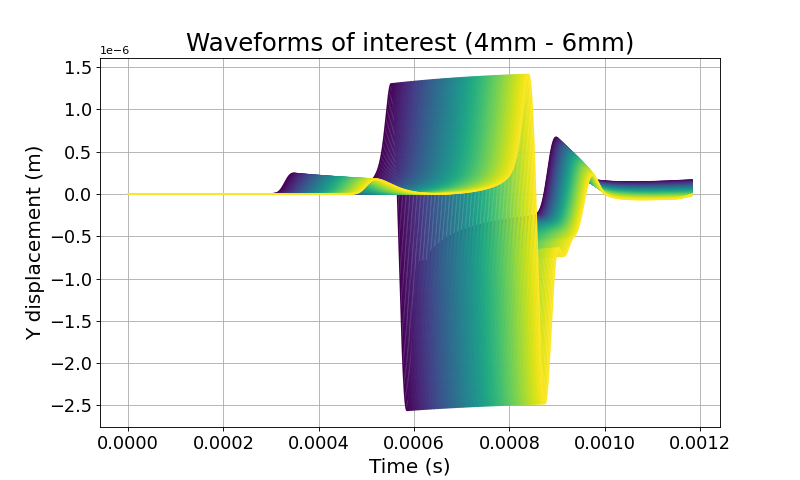}}\par\medskip
    
    \caption{Graph of the vertical displacement for (a) a node 4.8 mm away from the impact and (b) all nodes in the 4 mm - 6 mm region of interest. neo-Hookean material with a Young's modulus of 175 kPa, a density of 1,116 kg $\text{m}^{-3}$ and a pre-stretch of 1.2 (20\% extension).}
    \label{Typical Waveform Shape}
\end{figure}

A typical y-displacement vs time graph 4.8mm away from the applied perturbation is given in Figure \ref{Typical Waveform Shape:a}. From these curves, important information about the speed of the elastic waves can be extracted and used to predict the material properties of the block of skin. For example, in Figure \ref{Typical Waveform Shape} we see that a fast wave arrives just before 0.4 ms followed by a larger peak and corresponding trough occurring just before 0.6 ms. Finally, we can see a peak occurring around 0.9 ms which is the result of the first wave reflecting off the bottom of the skin before reaching the node. The first wave is the supersonic shear wave and the second (larger) wave is the Rayleigh surface wave (to be discussed further in Section \ref{FE Results Section}).

The deformed coordinates of the nodes, i.e. their position after the pre-stretch step, can be used in conjunction with the arrival time of the wave to determine the wave velocity. However, this naive approach makes the implicit assumption that the wave is traveling at constant speed. It also requires precise knowledge about the wave generation method. For example, if a pressure is applied, it is necessary to know the precise area that the pressure was applied to as well as the duration of the perturbation.

By measuring multiple waveforms at different distances from the perturbation, see Figure \ref{Typical Waveform Shape:b}, it is possible to extract more accurate measures of the wave speed while avoiding such assumptions. Finally, to avoid inherent discretisation errors when determining the arrival time of the ``maximum'' y-displacement, the waveform data points were interpolated by means of a quadratic spline function, as implemented in the Python function ``InterpolatedUnivariateSpline'' from the Python sub-package ``scipy.interpolate'' \cite{virtanen_scipy_2020}. This smooth spline function passes through all datapoints and can be differentiated to find a more accurate arrival time of the waveform peak. Therefore, in our study, the speeds of the Rayleigh and supersonic shear waves ($v_R$ and $v_s$) were taken to be the average of the wave speeds within the region of interest.

\subsection{Input Space Sampling} \label{Input Space Sampling Section}

Our goal was to train a ML model that could predict the material parameters of the skin using only the Rayleigh wave speed and supersonic shear wave speed described in Section \ref{FE Section}. In order to have a model that is capable of accurate predictions for a wide variety of subjects with different combinations of material parameters, it was necessary to sample carefully from the input space.

We had to explore a four-dimensional input space of $E$, $\beta$, $\rho$ and $\lambda_1$ (Young's modulus, beta, density and pre-stretch). To guarantee good coverage of this input space, a Latin hypercube sampling method \cite{mckay_comparison_1979} was employed. Specifically, the function ``LatinHypercube'' from the Python sub-package ``scipy.stats.qmc'' (Quasi-Monte Carlo) \cite{virtanen_scipy_2020} was used to generate 5,000 samples using a neo-Hookean material model ($\beta = -1$) and an additional 5,000 samples using a Mooney-Rivlin material model. The material parameter ranges were chosen specifically to closely resemble those of \textit{in vivo} human skin.

The neo-Hookean material model is hyperelastic and is commonly used to describe incompressible material response due to the simplicity of the form \cite{behera_peridynamic_2020}. In Abaqus, its strain energy function $U$ can be expressed as:
\begin{equation}
    U = C_{10} \left( \Bar{I}_1 - 3 \right) + \frac{1}{D_1} \left( J_{el} - 1 \right)^2,
\end{equation}
where $C_{10}$ and $D_1$ are material parameters, $\Bar{I}_1$ is the reduced first strain invariant of the left Cauchy-Green tensor and $J_{el}$ is the elastic volume strain \cite{smith_abaqusstandard_2009}. $C_{10}$ and $D_1$ can be related to the stiffness measured by the Young's modulus $E$ and the incompressibility measured by the Poisson ratio $\nu$ by:
\begin{align}
    C_{10} = \frac{E}{6},
    \qquad D_{1} = \frac{9 - 18 \nu}{E \left( 1 + \nu \right)}.
\end{align}

The Mooney-Rivlin material model can be viewed as an extension of the neo-Hookean form as it adds a term that depends on the reduced second strain invariant $\Bar{I}_2$ \cite{smith_abaqusstandard_2009}. In Abaqus, its strain energy function can be expressed as:
\begin{equation}
    U = C_{10} \left( \Bar{I}_1 - 3 \right) + C_{01} \left( \Bar{I}_2 - 3 \right) + \frac{1}{D_1} \left( J_{el} - 1 \right)^2,
\end{equation}
where $C_{10}$, $C_{01}$ and $D_1$ are material parameters and $J_{el}$ is the elastic volume strain. In this case, the initial Young's modulus $E$ is expressed as \cite{smith_abaqusstandard_2009}:
\begin{equation}
    E = 6 \left( C_{01} + C_{10} \right).
\end{equation}

Therefore, to express the model parameters $C_{10}$, $C_{01}$ and $D_1$ in terms of $E$ and $\nu$, a unitless parameter $\beta$ is introduced:
\begin{align}
    C_{10} = \frac{E}{12} \left( 1 - \beta \right),
    \qquad C_{01} = \frac{E}{12} \left( 1 + \beta \right),
    \qquad \nu = \frac{9 - 18 \nu}{E \left( 1 + \nu \right)},
\end{align}

where $\beta$ ranges from $-1$, representing a pure neo-Hookean material, to $+1$, representing a pure Mooney-Rivlin case. 

The density of skin $\rho$ is often assumed to be a fixed value, for example the value $1116 \, \text{kg}/\text{m}^3$ \cite{li_determining_2011}. However, to allow for some variation due to hydration and other factors, we allowed the value to vary by $\pm 5\%$. The skin's pre-stretch range in the principal direction $\lambda_1$ varies from study to study depending on the measurement procedure. Jor et al. reported a maximum skin retraction of approximately $40\%$ for porcine skin \cite{jor_estimating_2011}, Deroy et al. reported contractions in the $10\% - 30\%$ range for canine skin \cite{deroy_non-invasive_2017} and finally, Ní Annaidh et al. reported the mean failure strain of excised human skin to be $54\% \pm 17\%$ \cite{ni_annaidh_characterization_2012}. For our study, a pre-stretch in the range from 5\% to 35\% was chosen.

The stiffness of human skin as measured by the Young's modulus $E$ has been reported extensively using various methods. Liang and Boppart reported forearm skin to have $E \in [50, 150]$ kPa using optical coherence elastography \cite{x_liang_biomechanical_2010}, Li et al. reported values of the forearm dermis in the range $E \in [152.27, 286.68]$ kPa by measuring surface waves using optical coherence tomography \cite{li_determining_2011}, and Diridollou et al. reported forearm skin with $E \in [80, 260]$ kPa using a suction device \cite{diridollou_skin_2001}. As such, for our study we selected a reasonably broad range of values between $50$kPa and $300$kPa. It should be noted here that while the Young's modulus is widely reported, there is a significant spread in the literature, due to variations in both the location of the skin on the body and the method used to identify the Young's modulus. For instance, the review paper by Joodaki and Panzer \cite{joodaki_skin_2018} presents a summary of studies on the Young's modulus of whole skin, with measurements varying significantly. These variations range from $1.09$ kPa (forearm of young female) as reported by Bader and Bowker \cite{bader_mechanical_1983} using an indentation test method, to tens of thousands of kPa reported by Grahame and Holt \cite{grahame_influence_2009} using a suction device. Note that we used a fixed value for the Poisson ratio ($\nu = 0.495$) assuming all materials nearly incompressible \cite{sanders_torsional_1973} \cite{hendriks_numerical-experimental_2003} \cite{reihsner_two-dimensional_1995}. 

In summary, 5,000 unique neo-Hookean and 5,000 unique Mooney-Rivlin subjects were generated with a Latin hypercube sampling technique using the material parameter ranges in Table \ref{Material Property Ranges Table}. Two different material models were used to examine if the training and performance of the ML model were affected by the material model employed, i.e. the neo-Hookean and Mooney-Rivlin formulations, which are among the most common material models used for the breast \cite{teixeira_review_2023}. For each subject, a FE simulation consisting of a static pre-stretch followed by a dynamic wave propagation technique (see Section \ref{FE Section}) were performed, and the average Rayleigh and supersonic wave speeds were stored. This dataset was then used to train ML models of interest as described in the subsequent sections.

\begin{table}[ht]
    \centering
    \begin{tabular}{|c|c|l|} \hline 
         Material Property & Range & Units\\ \hline 
         $E$ & $\left[ 50, 300 \right]$ & kPa\\ 
         $\beta$ & $\left[ -1, 1 \right]$ & \\ \
         $\rho \,\,\,$ & $\left[ 1060.2, 1171.8 \right]$ & $ \text{kg}/\text{m}^3$\\ 
         $\lambda_1$ & $\left[ 1.05, 1.35 \right]$ & \\ \hline
    \end{tabular}
    \caption{Material property ranges used when sampling from the input space.}
    \label{Material Property Ranges Table}
\end{table}

\subsection{Statistical Emulation} \label{Statistical Emulation Section}

A ``simulator'' is a mathematical representation of a physical system implemented on a computer. Simulators are often deterministic (where the same set of inputs will always give the same output) and have been used to investigate real-world systems in many fields of research \cite{bastos_diagnostics_2009} \cite{sacks_design_1989} \cite{gramacy_surrogates_2020} \cite{santner_design_2018}. While simulators can be the gold standard for replication of complex real-world behaviour, they are often computationally expensive and thus may not be suitable for all applications. For example, the gold standard method for determining the complex mechanical response of skin or other biological materials is a FE model. However, in practice, a three-dimensional FE simulation could take hours to run, making this method infeasible in many clinical settings, where an analysis needs to be performed quickly. Similarly, if it were of interest to see how small changes to each individual FE input affected the output (i.e. performing sensitivity analysis \cite{razavi_future_2021}), it would be necessary to run many FE simulations, which could be very computationally expensive.

An ``emulator'' is a data-driven model that uses outputs from the simulator to reconstruct the simulation in a relatively computationally inexpensive manner. The process of constructing an emulator involves the creation of a training dataset, which is constituted of evaluations of the simulator for diverse input parameters. This dataset is subsequently used to train the emulator, enabling it to predict simulation outputs for input instances that have not been directly evaluated within the simulator \cite{bastos_diagnostics_2009} \cite{gramacy_surrogates_2020} \cite{santner_design_2018}. These cheap, fast, accurate approximations of the true simulator outputs can allow for clinical/real-world use, sensitivity analysis, efficient optimisation, uncertainty quantification, etc.

In our study, the simulator is the two-dimensional FE model described in Section \ref{FE Section}, which has the material parameters of the skin as its inputs ($E$, $\rho$, $\lambda_1$ and $\beta$), and its outputs are the average Rayleigh and supersonic shear wave speeds ($v_{R}$ and $v_{s}$) in the 4-6 mm region of interest; see Figure \ref{Schematic of Emulator and ML Model}. The training dataset for the emulator consists of the 10,000 input instances obtained from the Latin hypercube design for the various combinations of the input parameters ($E$, $\rho$, $\lambda_1$ and $\beta$), defined in Section \ref{Input Space Sampling Section}, and the associated simulation velocity outputs ($v_{R}$ and $v_{s}$). According to the value of $\beta$, the dataset is comprised of 5,000 instances of neo-Hookean subjects and 5,000 Mooney-Rivlin subjects. While in principle many statistical and ML models can be employed as emulators (regression models, Gaussian process, random forests, support vector machines, neural networks etc.) \cite{villa-vialaneix_comparison_2012} \cite{zhang_estimation_2021} \cite{bhosekar_advances_2018}, here we consider Gaussian process (GP) \cite{rasmussen_gaussian_2004}. Specifically, a Gaussian process regression model as implemented in ``GaussianProcessRegressor'' in the Python sub-package ``sklearn.gaussian\_process'' was used \cite{pedregosa_scikit-learn_2011}. The GP regression model was trained using a radial-basis function (RBF) kernel with length scale 1.0 as implemented in ``RBF'' in the Python sub-package ``sklearn.gaussian\_process.kernels'' \cite{pedregosa_scikit-learn_2011}. It was found to provide high predictive performance and it also allows for uncertainty quantification by giving the mean and standard deviation as output when predicting, see Section \ref{Statistical Emulation Results Section}.

\subsection{Non-invasive Prediction of Material Properties} \label{Prediction of Material Properties Section}

As discussed in Section \ref{Introduction}, there is a need to be able to measure the material properties of \textit{in vivo} skin with a non-invasive procedure. In this study, we suggest that the speed of a surface wave traveling through the skin contains information about the material properties, which can be extracted. In theory, determining the wave speeds could be very straightforward, see Section \ref{Experimental Validation Section}. However, using these wave speeds to quantify stress and the pre-stretch \textit{in vivo} requires solving a complex inverse problem in real time.

\begin{figure}[ht]
  \centering
  \includegraphics[width=0.9\textwidth]{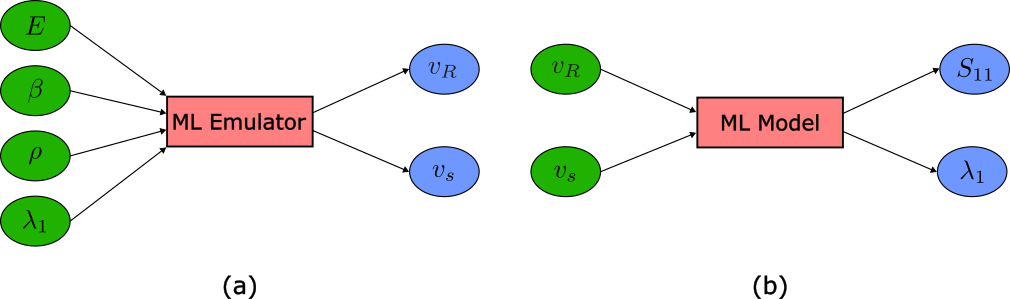}%
  \caption{Schematic of (a) the ML emulator used to relate the four input variables $E$, $\beta$, $\rho$ and $\lambda_1$ (the Young's modulus, beta parameter, density and pre-stretch respectively), to the two output variables $v_R$ and $v_s$ (the Rayleigh wave speed and supersonic shear wave speed, respectively). (b) The ML model used to infer the steady-state stress, $S_{11}$, in the principal directions of stretch and the natural pre-stretch of the subject's skin, $\lambda_1$, from the input variables corresponding to the Rayleigh wave speed $v_R$ and the supersonic shear wave speed $v_s$.}
  \label{Schematic of Emulator and ML Model}
\end{figure}

Similarly to the use of an emulator to reproduce the outputs of a computer simulation, we propose using a ML model as a computationally efficient way of solving the inverse problem of inferring stress and pre-stretch from the wave speeds. For this model, the velocity of the supersonic shear wave $v_s$ and the Rayleigh wave speed $v_R$ are now the input variables. The two target variables are the steady-state (after the pre-stretch but before the wave propagation) stress in the principal direction of stretch $S_{11}$ and the natural pre-stretch of the subject’s skin $\lambda_1$; see Figure \ref{Schematic of Emulator and ML Model}. These target variables were chosen as they are independent of the material model being used (unlike the Young's modulus $E$ for example) and are the parameters of most interest in a surgical setting. The same dataset described in Section \ref{Statistical Emulation Section} obtained from the simulator is used for training, where in this case the inputs are $(v_R, v_s)$ and the targets are $(S_{11}, \lambda_1)$.

Again, while in principle this model could be of any form, in this case we consider a Gaussian process regression model as it provided high predictive performance and also allowed for uncertainty quantification; see Section \ref{Material Property Prediction Results Section}.

\subsection{Experimental Validation} \label{Experimental Validation Section}

\begin{figure}[ht]
  \centering
  \includegraphics[width=\textwidth]{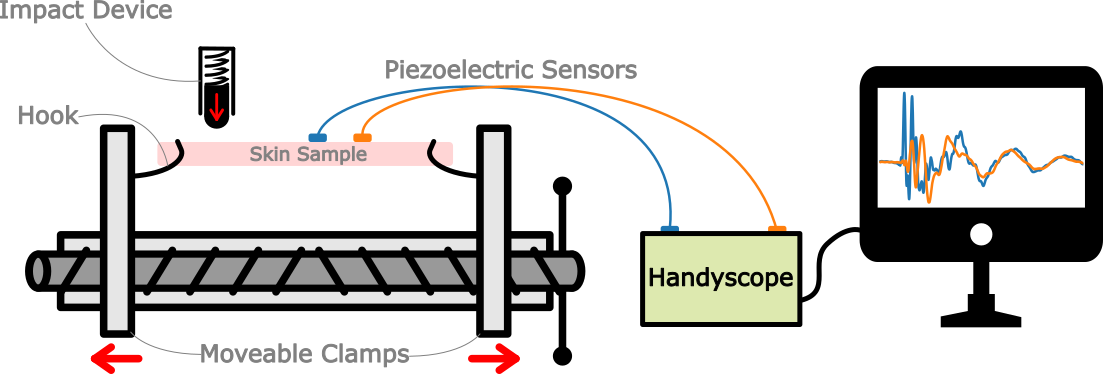}%
  \caption{Schematic of the experimental device used to collect wave speed measurements. The device consists of a synthetic tissue (Simulab) which is stretched uniaxially to a known pre-stretch value. A spring-loaded device provides the perturbation and two piezoelectric sensors a known distance apart record the shape of the waveform for analysis.}
  \label{Experimental Set Up}
\end{figure}

To demonstrate that surface wave speed data of the type described in Section \ref{FE Section} can be collected easily and cheaply, a custom device was created consisting of two piezoelectric sensors and a custom uniaxial stretching apparatus, see Figure \ref{Experimental Set Up}.

The skin sample used was a synthetic tissue from Simulab (Seattle, USA) and comprised of a single layer of thickness 1 mm \cite{kho_mechanical_2023}, designed to replicate the epidermis of human skin. The uniaxial stretching apparatus consisted of a fixed base and two moving clamps mounted on a leadscrew. The synthetic tissue was cut into strips with known dimensions $50 \, \text{mm} \times 24 \, \text{mm}$ and secured to the clamps using a number of hooks that pierced through the skin. A known uniaxial pre-stretch could then be applied to the skin by rotating the leadscrew.

\begin{figure}[ht]
  \centering
  \includegraphics[width=0.8\textwidth]{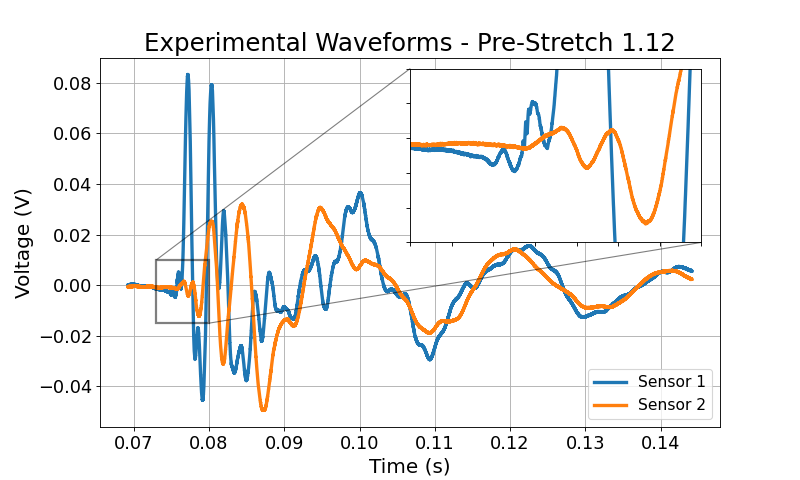}%
  \caption{Typical graph of Voltage vs Time from the two piezoelectric sensors. Synthetic tissue (Simulab) with a pre-stretch value of 1.12 (i.e. 12\% stretch), the distance between the sensors was 17.13 mm. The most consistent results were found when comparing the initial voltage deviations when the wave initially arrives at the sensors, shown in inset plot.}
  \label{Experimental Waveforms}
\end{figure}

A spring-loaded device was installed at a fixed distance above the skin sample, capable of providing repeatable perturbations normal to the surface of the skin, to generate a surface wave. To measure the shape of the waveform, two piezoelectric sensors (TE Connectivity Measurement Specialties) were fixed perpendicular to the direction of the uniaxial stretch, placed at a known distance apart and at a known distance from the impact site. The two sensors were connected to a Handyscope HS3-100 oscilloscope (TiePie Engineering), from which it was possible to visualise the voltage generated by the sensor as a function of time, see Figure \ref{Experimental Waveforms}.

Note that in Figure \ref{Experimental Waveforms} we see more oscillatory behaviour than in the FE model data (Figure \ref{Typical Waveform Shape}). This is likely due to wave reflection off the bottom of the skin sample and the inherently more complex surface wave propagation behaviour we would expect in three dimensions in a synthetic tissue sample. It should also be noted that it was only possible to extract information about the Rayleigh wave as the supersonic wave was not visible. This may be due to wave attenuation or its smaller amplitude (see Figure \ref{Typical Waveform Shape}), making it more difficult to detect.

Using this setup, it was possible to experimentally determine the effect of pre-stretch on the surface wave velocity. Four pre-stretch values of 1.12, 1.19, 1.22 and 1.27 were considered and five measurements were performed for each pre-stretch value. Note that the voltage-time waveforms obtained experimentally are analogous to the displacement-time graphs obtained from the FE simulations. Surface wave speed values were obtained by comparing the ``arrival time'' of particular features of the two waveforms and using the known distance between the piezoelectric sensors.

\section{Results}

\subsection{Finite Element Results} \label{FE Results Section}

\begin{figure}[ht]
  \centering
  \includegraphics[width=0.7\textwidth]{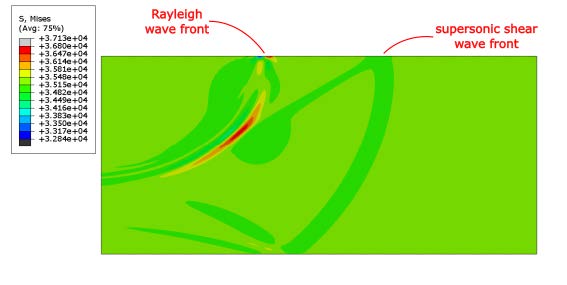}%
  \caption{Von Mises stress (Pa) in the deformed neo-Hookean material with a Young's modulus of 175 kPa, a density of 1,116 kg $\text{m}^{-3}$ and a pre-stretch of 1.2 (20\% extension). This frame of the simulation was taken after 0.553 ms of wave propagation. The (larger) Rayleigh wave can be seen traveling along the surface of the skin as well as the faster supersonic shear wave.}
  \label{Dynamic Wave Propagation}
\end{figure}

As discussed in Section \ref{FE Section}, the first step of each simulation was to perform a uniaxial static pre-stretch to simulate a pre-stressed ``\textit{in vivo}'' state before the wave was propagated along the surface of the skin. After the pre-stretch, a motion was generated through the skin, see Section \ref{FE Section}. Both the Rayleigh and supersonic shear waves can be seen propagating through the material in Figure \ref{Dynamic Wave Propagation}.

\begin{figure}[ht]
  \centering
  \includegraphics[width=0.7\textwidth]{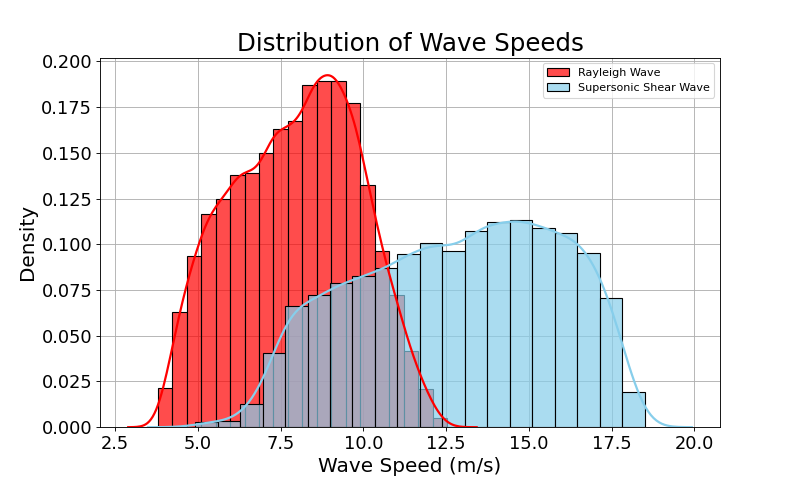}%
  \caption{Histograms of the distribution of the Rayleigh and supersonic shear wave speeds for all 10,000 subjects.}
  \label{Wave Speed Distribution}
\end{figure}

After running all 10,000 simulations, it was found that the Rayleigh wave speed traveled between 3.78 and 12.53 m/s while the supersonic shear wave traveled between 4.9 and 18.5 m/s, with the supersonic shear wave always traveling faster than the Rayleigh wave. The distributions of the wave speeds are shown in Figure \ref{Wave Speed Distribution}. It should be noted here that the Rayleigh wave speeds from the FE simulation are in very good agreement with the predicted analytical results from Equation \ref{Modified Flavin Equation} ($R^2 = 0.9951$). The analytical wave speeds are only slightly faster than the FE wave speeds, within $1.59\%$ on average.

\subsection{Statistical Emulation Results} \label{Statistical Emulation Results Section}

The exact performance of any ML model is dependent on the train-test split of the data, i.e. which data points are used to train the model and which data points are withheld to test the performance of the trained model. So, to provide a fair assessment of the ability of the emulator to reproduce the simulator outputs, a 10-fold cross-validation procedure is implemented \cite{hastie_elements_2009}. In the procedure, the dataset is randomly split into 10 folds of 1,000 subjects. In turn, each fold is used as a test set, while the other nine folds are employed to train the Gaussian process regression model. The mean of the $R^2$ \cite{lewis-beck_applied_2015} computed between estimated and simulation-outputted Rayleigh and supersonic velocities is employed as a metric to assess the predictive performance. This $R^2$ is averaged across the 10 fold replications, giving an average performance of $0.9993 \pm 0.0003$. This result indicates that the emulator is able to reproduce the simulation outputs in a stable manner to a very high degree of accuracy.

\begin{figure}[ht]
    \centering
    \subfloat[]{\label{GP Emulator Performance:a}\includegraphics[width=.5\linewidth]{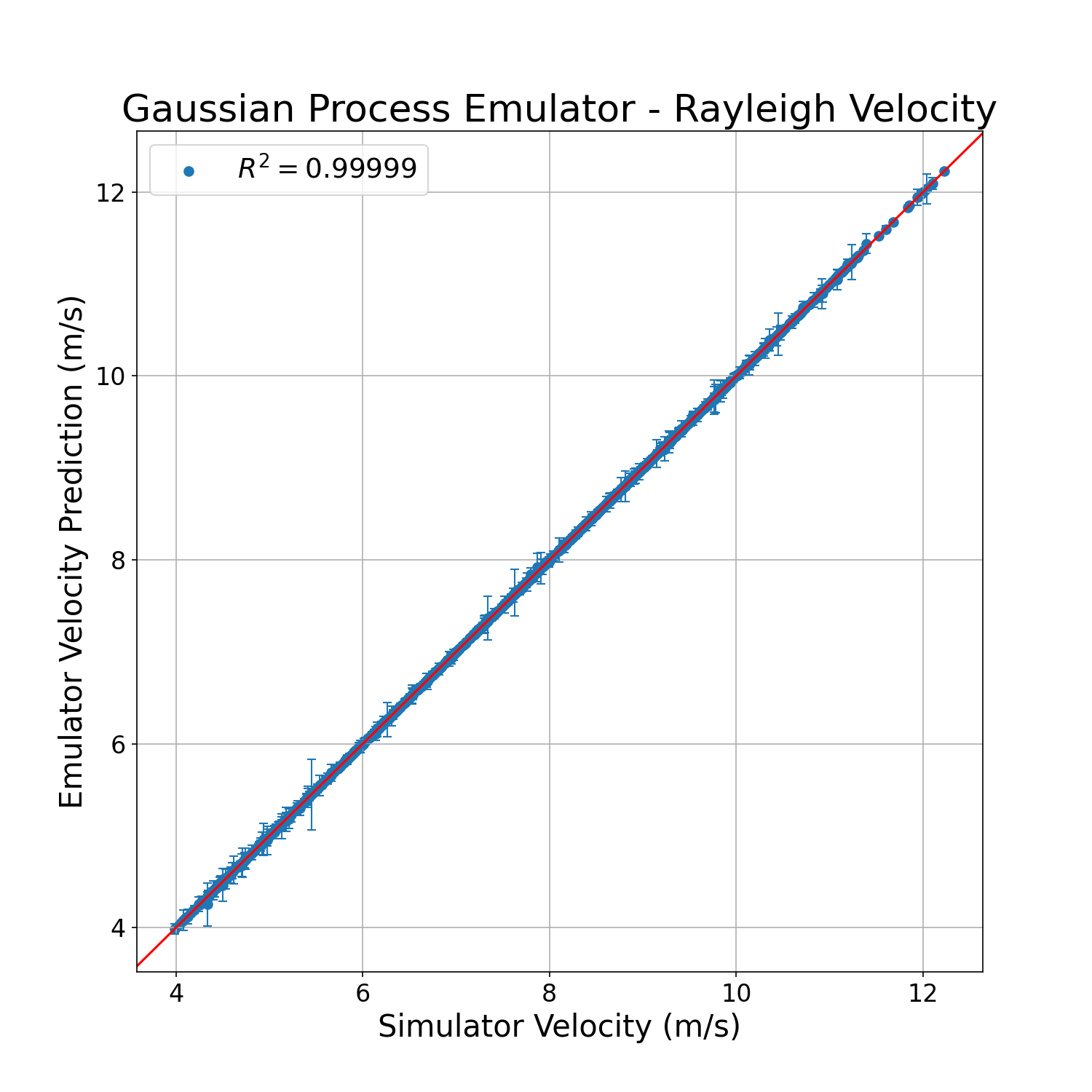}}
    \subfloat[]{\label{GP Emulator Performance:b}\includegraphics[width=.5\linewidth]{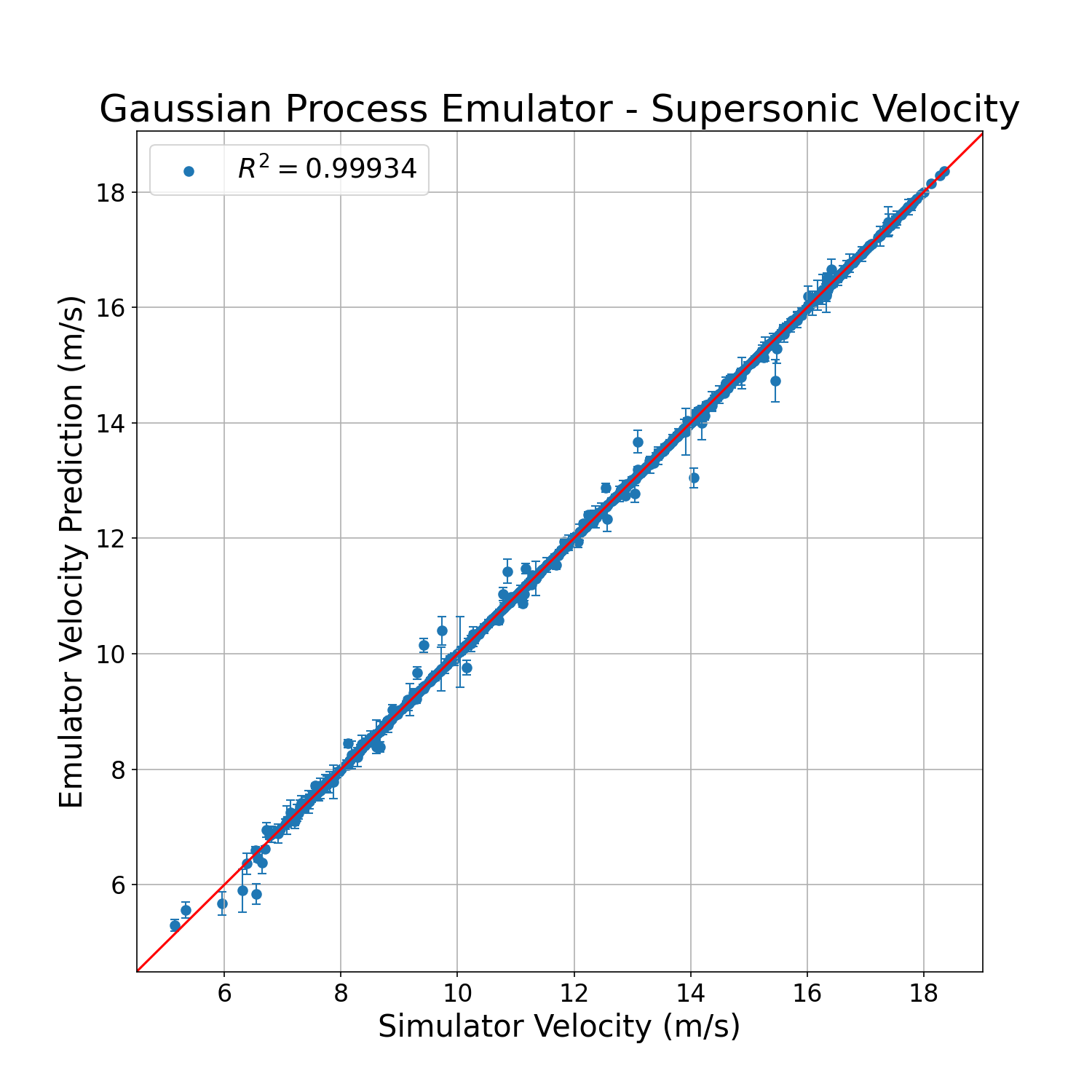}}\par\medskip
    
    \caption{Performance of the multi-output Gaussian process regression emulator trained on 90\% of the dataset and tested on the remaining unseen 10\%. For each datapoint, the x coordinate is the ``true'' wave speed extracted from the FE simulation, the corresponding y coordinate is the GP prediction of the wave speed given the set of inputs $E$, $\beta$, $\rho$ and $\lambda_1$ for that subject; the error bar is the 99\% credible interval for the GP prediction. As shown, the emulator has extremely high predictive power.}
    \label{GP Emulator Performance}
\end{figure}

To get a visual indication of the predictive performance we can consider one such 90\%/10\% train-test split, i.e. where one fold of the data is withheld as an unseen test set and the GP regression model is trained on the remaining nine folds. By comparing the simulation outputs to the predictions from the emulator for the unseen test set data points we see that the emulator model is capable of reproducing very similar outputs to the simulator at a greatly reduced computational cost; see Figure \ref{GP Emulator Performance}.

Once trained, our emulator can be used to make predictions of Rayleigh and supersonic wave speeds for unknown sets of inputs in a fraction of the time it would take to run the full simulator. For example, a typical run of our FE simulation on a single CPU would take approximately 6 minutes (excluding data extraction and post processing time). By contrast, a new prediction from the GP emulator takes approximately 30 milliseconds, a reduction in complexity of 4 orders of magnitude. With such a reduction in computing time, using the emulator it is possible to perform a sensitivity analysis of the velocities as a function of ranges of the input parameters.

First, a large dataset was generated using the emulator, 20 equally spaced points were taken from the range of each input variable ($E$, $\rho$, $\beta$, $\lambda_1$), and each possible combination of points was considered, giving a dataset of 160 thousand observations. Following the method of sensitivity analysis from Section 7.2.2 of The Design and Analysis of Computer Experiments \cite{santner_design_2018}, two first order regression models were fitted:
\begin{align} 
    v_R^* &= \alpha_0^* + \alpha_1^* E^* + \alpha_2^* \rho^* + \alpha_3^* \beta^* + \alpha_4^* \lambda_1^*, \label{sens analysis regression model 1 equation}\\
    v_s^* &= \alpha_0^* + \alpha_1^* E^* + \alpha_2^* \rho^* + \alpha_3^* \beta^* + \alpha_4^* \lambda_1^*, \label{sens analysis regression model 2 equation}
\end{align}
where $x^*$ denotes standardised value of $x$, given by:
\begin{equation*}
    x^* = \frac{x - \bar{x}}{\sigma},
\end{equation*}
where $\bar{x}$ is the mean of $x$ and $\sigma$ is the standard deviation of $x$.

The regression coefficients $\alpha_i^*$ in Equations \ref{sens analysis regression model 1 equation} and \ref{sens analysis regression model 2 equation} are known as the standardised regression coefficients (SRCs). For example $\alpha_1^*$ measures the change in our target variable ($v_R^*$ or $v_s^*$) due to a unit standard deviation change in our input $E$. Because all variables are on a common scale after standardisation, the magnitude of the estimated SRCs tells us the relative sensitivity of the output to each input. The output is most sensitive to the input that has the largest absolute SRC value \cite{santner_design_2018}. Table \ref{sens analysis regression coefficients table} shows the computed SRCs where we see that the target variables are most sensitive to changes in the Young's modulus $E$ and are least sensitive to changes in the density $\rho$. This result is consistent with Equation \ref{Modified Flavin Equation} where we see that the Rayleigh wave speed is directly proportional to $\sqrt{E/\rho}$. Note that when sampling from the input space (see Section \ref{Input Space Sampling Section}) we allowed $E$ to have a much larger variation than $\rho$, which, in the literature, is often taken to be constant. Interestingly, we also see that the Rayleigh wave speed is more sensitive to $\lambda_1$ and less sensitive to $\beta$ while conversely, the supersonic wave speed is more sensitive to $\beta$ and less sensitive to $\lambda_1$.

\begin{table}[ht]
    \centering
    \begin{tabular}{|c|c|c|} \hline
        Input Variable & Estimated $\alpha_i^*$ ($v_R$) & Estimated $\alpha_i^*$ ($v_s$)\\ \hline
        Young's Modulus ($E$) & 0.9447 & 0.9328 \\
        Density ($\rho$) & -0.0609 & -0.0621 \\
        Beta ($\beta$) & -0.1229 & -0.2493 \\
        Pre-Stretch ($\lambda_1$) & 0.2630 & -0.1359 \\ \hline
    \end{tabular}
    \caption{Standardised regression coefficients for Equations \ref{sens analysis regression model 1 equation} and \ref{sens analysis regression model 2 equation}. Coefficients indicate that the velocity outputs are most sensitive to the Young's modulus $E$ and are least sensitive to the density $\rho$. Note that this analysis is likely to be reasonable because the $R^2$ associated with the fitted models are 0.9796 and 0.9534 for $v_R$ and $v_s$, respectively \cite{santner_design_2018}.}
    \label{sens analysis regression coefficients table}
\end{table}

\begin{figure}[htb!]
    \centering
    \subfloat[]{\label{GP Sens Analysis:a}\includegraphics[width=.5\linewidth]{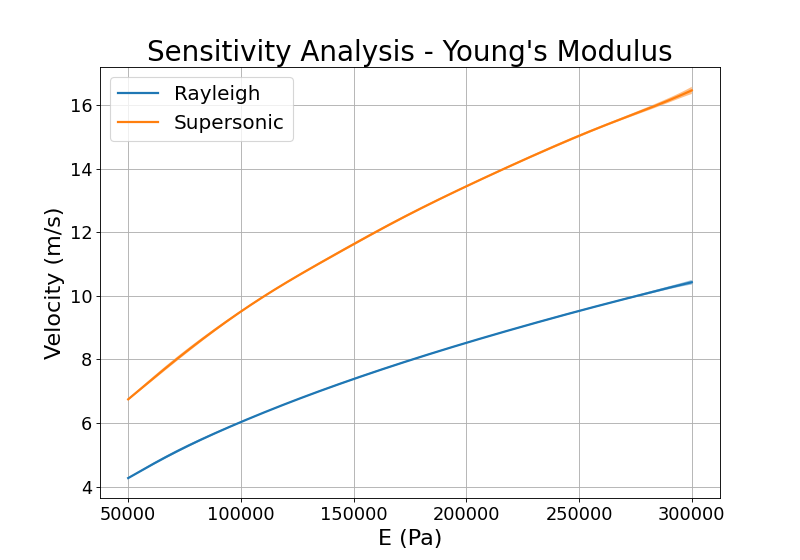}}
    \subfloat[]{\label{GP Sens Analysis:b}\includegraphics[width=.5\linewidth]{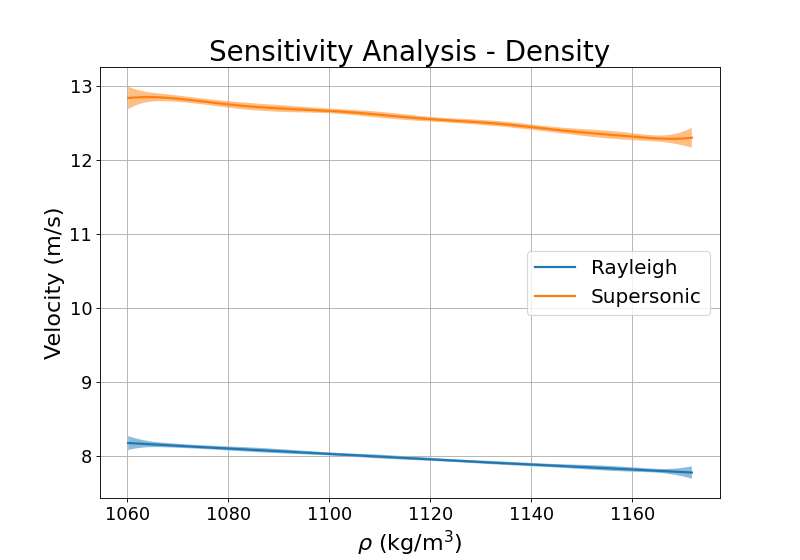}}\\
    \subfloat[]{\label{GP Sens Analysis:c}\includegraphics[width=.5\linewidth]{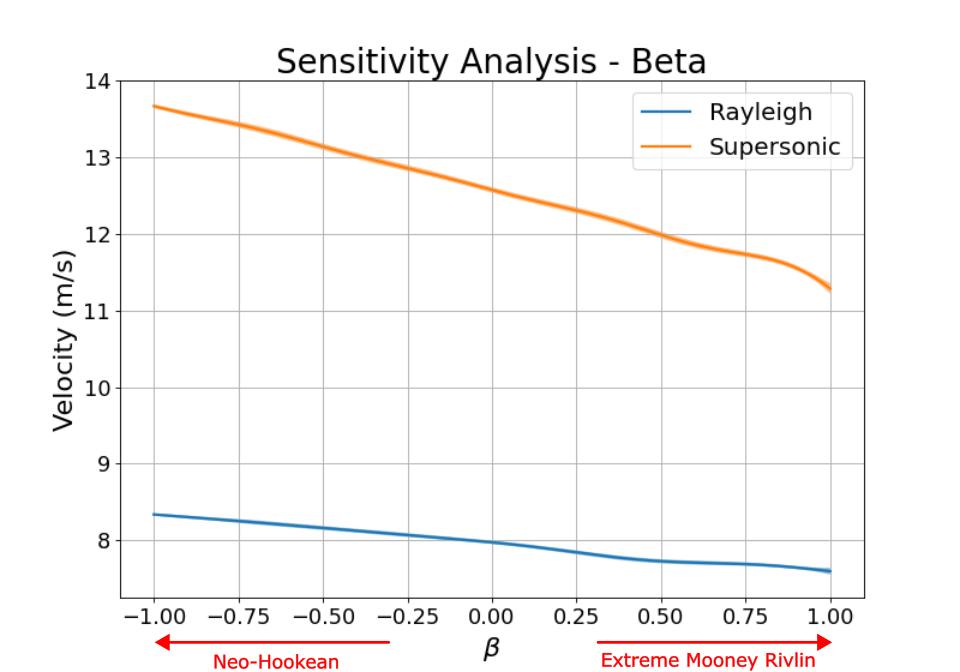}}
    \subfloat[]{\label{GP Sens Analysis:d}\includegraphics[width=.5\linewidth]{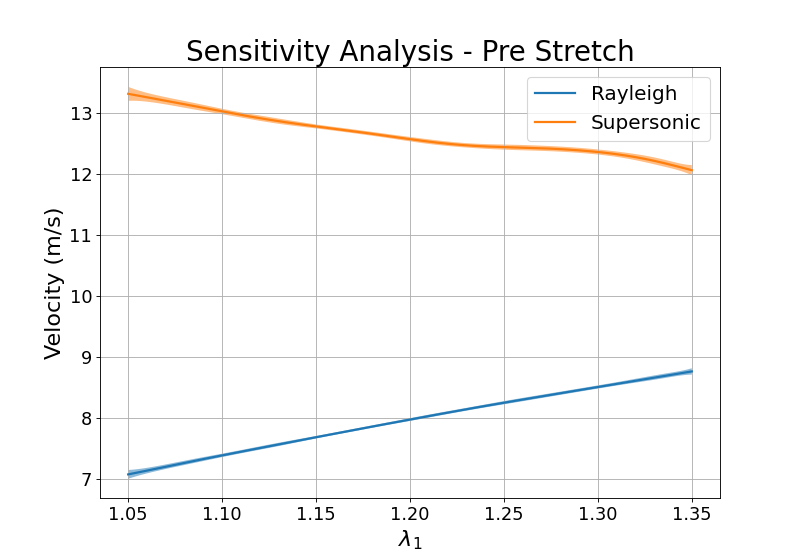}}\par\medskip 
    
    \caption{Results of the sensitivity analysis where each input (a) $E$, (b) $\rho$, (c) $\beta$ and (d) $\lambda_1$ was allowed to vary along their full range while each of the other input variables were held at their mean values, $E = 175$ kPa, $\beta = 0$, $\rho = 1,116$ kg $\text{m}^{-3}$ and $\lambda_1 = 1.2$ (20\% extension), and the effect on the output variables $v_R$ and $v_s$ was observed. Shaded region is the 99\% credible interval for the GP prediction.}
    \label{GP Sens Analysis}
\end{figure}

A visual representation of the effect that each input variable has on the response wave speed can be obtained through conditional plots where all input are fixed at their mean and one variable is allowed to vary in its full range of values, see Figure \ref{GP Sens Analysis}. The figures show that when the other material parameters are fixed at their mean and the Young's modulus is increased, both the Rayleigh and supersonic velocities also increase. Conversely, an increase in the density causes the wave speeds to decrease slightly (due to the relatively small amount of variance in density that was sampled). Interestingly, we see that there is a significant decrease in the supersonic wave speed in the transistion from a pure neo-Hookean material ($\beta = -1$) to an extreme Mooney-Rivlin material ($\beta = +1$), whereas there is a much weaker decrease in the Rayleigh wave speed. We can also see that, as expected, the additional stretch causes the Rayleigh wave speed to increase significantly, but surprisingly, it causes the supersonic shear wave speed to decrease. This phenomenon has been demonstrated experimentally in a recent publication by Li et al., who performed uniaxial extension on a rubber sample at different magnitudes of pre-stress and observed the variations of the phase velocities of the Rayleigh and supershear waves \cite{li_supershear_2022}.

While our emulator is performing the same task as Equation \ref{Modified Flavin Equation}, the ML method used is non-parametric, and as such, it is not possible to directly extract a simple equation which relates the material parameter inputs to the output velocities. However, it is possible to interpolate the complex relationship between our input variables ($E$, $\rho$, $\beta$ and $\lambda_1$) and our output variables ($v_R$ and $v_s$) that was learned by the Gaussian process emulator. The interpolation was performed using a parametric linear regression model. Specifically, the model ``LinearRegression'' from the Python sub-package ``sklearn.linear\_model'' \cite{pedregosa_scikit-learn_2011} was fit to the large dataset generated by the emulator during the sensitivity analysis, allowing us to extract the following equations:

\begin{align}
    \begin{split}
        v_R = &-0.1519 - (3.8577\times 10^{-11})E^2 + (3.7251\times 10^{-5})E\\&- (3.4424\times 10^{-3})\rho - 0.3868 \beta + 5.5120 \lambda_1 \label{interpolated equation for Rayleigh velocity}
    \end{split}\\
    \begin{split}
        v_s = &15.8983 - (6.0573\times 10^{-11})E^2 + (5.8806\times 10^{-5})E\\&- (5.6327\times 10^{-3})\rho - 1.2569 \beta - 4.5659 \lambda_1  \label{interpolated equation for supersonic velocity}  
    \end{split}
\end{align}

Note that Equations \ref{interpolated equation for Rayleigh velocity} and \ref{interpolated equation for supersonic velocity} are simplifications of the true complex relationship between our variables. They are data-driven equations (not physics-derived, like Equation \ref{Modified Flavin Equation}) that are conditional on the emulator model chosen and the FE model used to generate the simulator dataset. The $R^2$ values for the interpolation of Equation \ref{interpolated equation for Rayleigh velocity} and \ref{interpolated equation for supersonic velocity} are 0.9905 and 0.9637, respectively. This means that the simplified parametric equations are a good approximation of the non-parametric ML model.

\subsection{Material Property Prediction Results} \label{Material Property Prediction Results Section}

As described in Section \ref{Prediction of Material Properties Section}, a GP regression model was employed to predict steady state stress $S_{11}$ in the principal direction of stretch and natural pre-stretch $\lambda_1$ of the subject's skin using the Rayleigh speed $v_R$ and the supersonic speed $v_s$ as inputs. 

As in Section \ref{Statistical Emulation Results Section}, we get a fair assessment of the ability of the GP regression model to predict the steady state stress $S_{11}$ and the natural pre-stretch $\lambda_1$ by performing a 10-fold cross-validation procedure. The mean of the $R^2$ values computed for estimated and simulation-outputted steady-state stress $S_{11}$ and natural pre-stretch $\lambda_1$ was employed as a metric to assess the predictive performance. This $R^2$ is averaged across the 10 fold replications, giving an average performance of $0.9570 \pm 0.0025$ when compared with the ``true'' values extracted from the FE simulations. From this, we can be confident that the predictive performance of the model is very strong and is not dependent on the initial split of the training and testing data.

\begin{figure}[htb]
    \centering
    \subfloat[]{\label{GP Model Performance:a}\includegraphics[width=.5\linewidth]{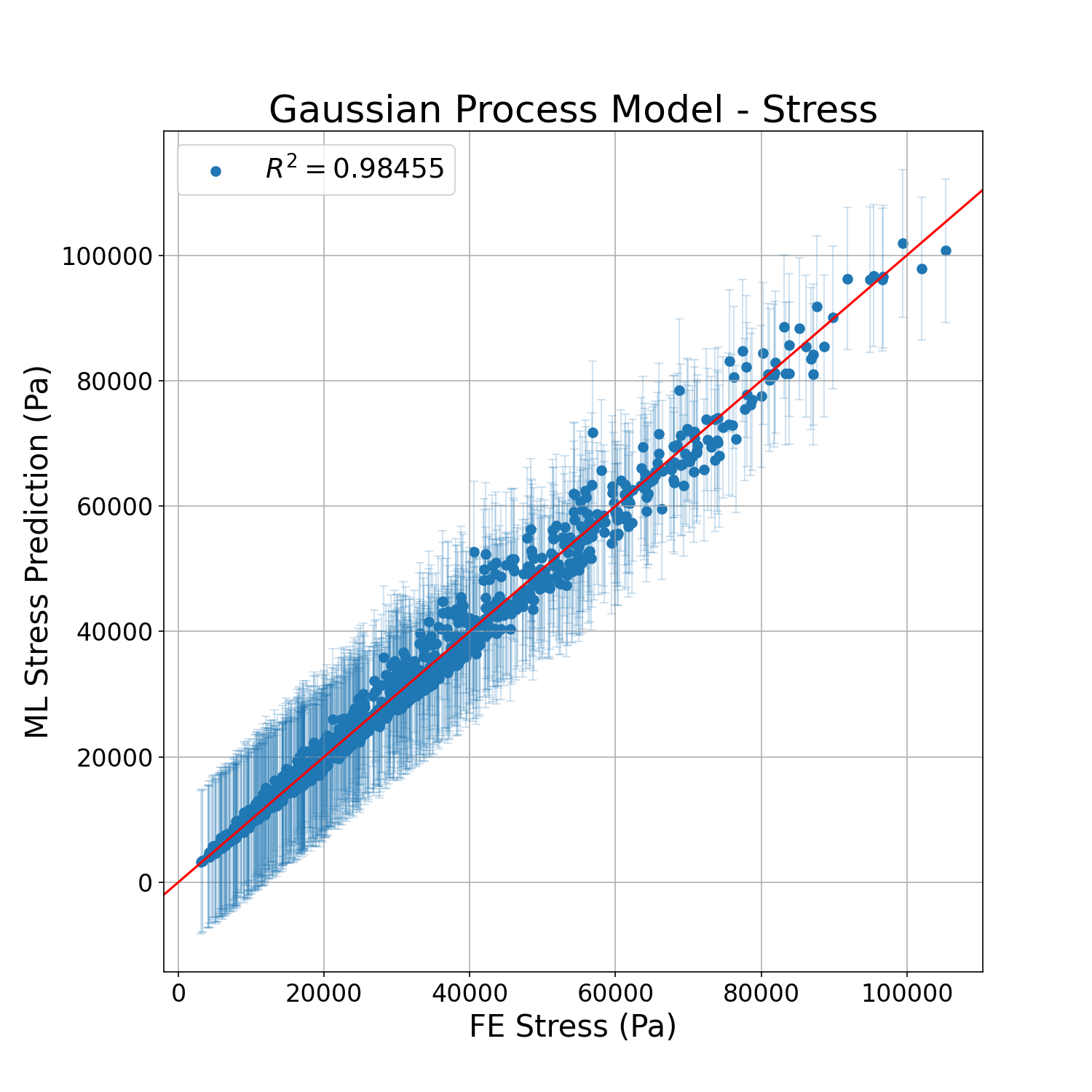}}
    \subfloat[]{\label{GP Model Performance:b}\includegraphics[width=.5\linewidth]{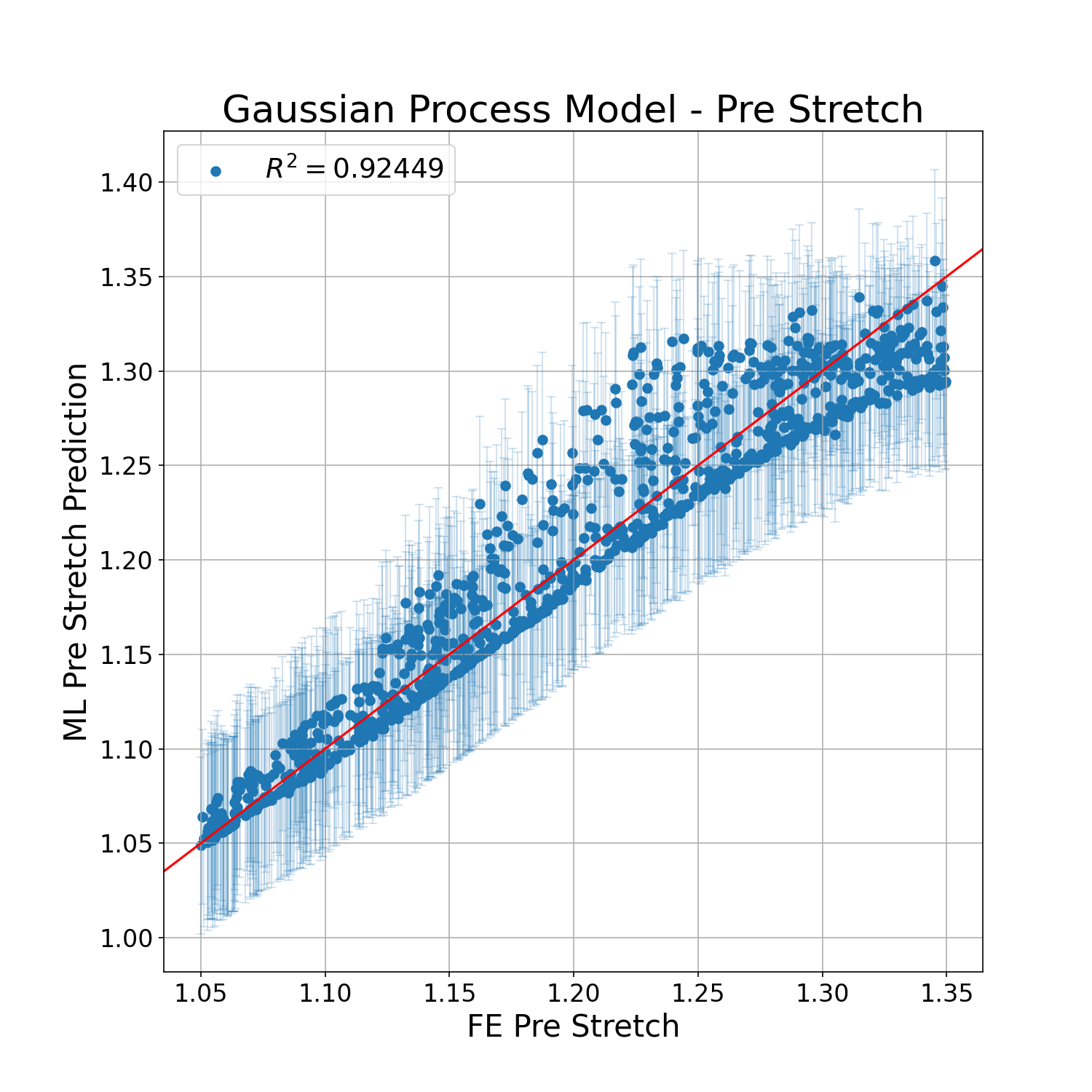}}\par\medskip
    
    \caption{Performance of the multi-output GP regression model trained on 90\% of the dataset and tested on the remaining unseen 10\%. For each datapoint, the x coordinate is the ``true'' stress/pre-stretch extracted from the FE simulation and the corresponding y coordinate is the GP prediction of the stress/pre-stretch given the wave speeds $v_s$ and $v_R$. The error bar is the 99\% credible interval for the GP prediction. As shown, the model has very strong predictive power for both output variables.}
    \label{GP Model Performance}
\end{figure}

Again, we get a visual indication of predictive performance by considering one such 90\%/10\% train-test split, i.e. where one fold of the data is withheld as an unseen test set and the GP regression model is trained on the remaining nine folds. By comparing the ``true values'' extracted from the FE simulations and the predictions from the GP model for the unseen test set data points, we see that the GP model is capable of accurate predictions of the stress and the pre-stretch of the skin using only the Rayleigh and supersonic wave speeds; see Figure \ref{GP Model Performance}. Unsurprisingly, as the stress/stiffness in the material has the largest influence on the wave velocities (see Figure \ref{GP Sens Analysis}), the model performs better at predicting the stress/stiffness, leading to a higher $R^2$ value. Interestingly, however, as the pre-stretch of the material has opposite effects on the Rayleigh and supersonic wave speeds (see Figure \ref{GP Sens Analysis:d}), the inclusion of the supersonic shear wave speed as an input to the GP model also allows accurate predictions of the pre-stretch. For comparison, the same GP model trained with the same train-test split with just a single input $v_R$ is still capable of making reasonably accurate predictions of $S_{11}$ ($R^2 = 0.8136$) but is incapable of accurate predictions of $\lambda_1$ ($R^2 = 0.1405$). We can also confirm this by performing the same sensitivity analysis procedure explained in Section \ref{Statistical Emulation Results Section} using the 10,000 measurements of $v_R$ and $v_s$ from the training dataset as our inputs. We see from the SRCs in Table \ref{sens analysis 2 regression coefficients table} that while the Rayleigh wave speed is the most important, the supersonic wave speed has a noticeable level of importance in the GP model.

\begin{table}[h]
    \centering
    \begin{tabular}{|c|c|c|} \hline
        Input Variable & Estimated $\alpha_i^*$ ($S_{11}$) & Estimated $\alpha_i^*$ ($\lambda_1$)\\ \hline
        Rayleigh wave speed ($v_R$) & 1.8920 & 2.3758 \\
        Supersonic wave speed ($v_s$) & -1.1168 & -2.2414 \\ \hline
    \end{tabular}
    \caption{Standardised regression coefficients for $S_{11}$ and $\lambda_1$. Coefficients indicate that $S_{11}$ and $\lambda_1$  are sensitive to both $v_R$ and $v_s$.}
    \label{sens analysis 2 regression coefficients table}
\end{table}

Again, we use our pre-trained GP model and our 10,000 combinations of $v_R$ and $v_s$ from the training data to extract a simplified parametric equation by fitting a linear regression model to interpolate the complex relationship between our input variables ($v_R$ and $v_s$) and our output variables ($S_{11}$ and $\lambda_1$):

\begin{align}
    \begin{split}
        S_{11} = &7639.4731 - 1163.2769 {v_R}^2 - 3983.7169 v_R\\&- 1388.9624 {v_s}^2 + 980.2333 v_s + 3394.7445 v_R v_s \label{interpolated equation for stress}
    \end{split}\\
    \begin{split}
        \lambda_1 = &1.1925 - 0.0306 {v_R}^2 + 0.1839 v_R\\& - 0.0084 {v_s}^2 - 0.1150 v_s + 0.0329 v_R v_s  \label{interpolated equation for pre-stretch}  
    \end{split}
\end{align}

Again, note that Equations \ref{interpolated equation for stress} and \ref{interpolated equation for pre-stretch} are simplified data-driven approximations of the true complex relationship. As such, Equations \ref{interpolated equation for stress} and \ref{interpolated equation for pre-stretch} are conditional on the emulator model chosen and the FE model used to generate the simulator dataset. The $R^2$ values for the interpolation of Equations \ref{interpolated equation for stress} and \ref{interpolated equation for pre-stretch} are 0.9973 and 0.9540, respectively.

\subsection{Experimental Validation Results} \label{Experimental Validation Results Section}

\begin{figure}[ht]
  \centering
  \includegraphics[width=0.8\textwidth]{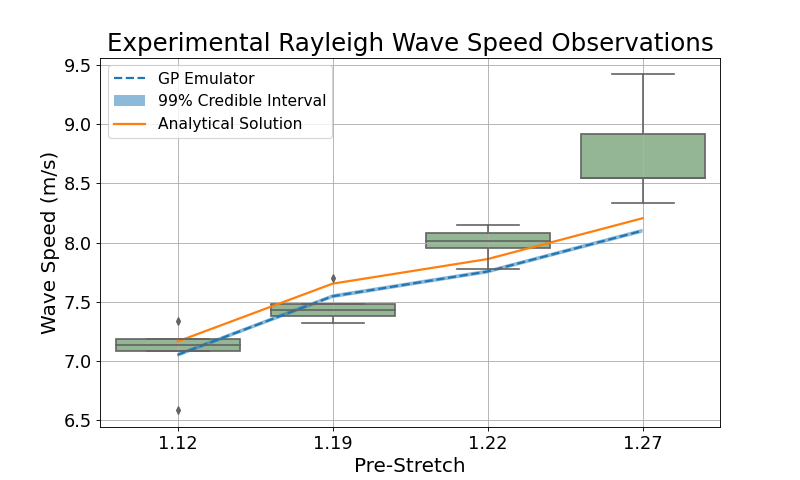}%
  \caption{Boxplots of the wave speeds measured experimentally for four different levels of pre-stretch. The GP emulator wave speed predictions and the analytical solution using Equation \ref{Modified Flavin Equation} are in blue and orange respectively, where the synthetic tissue was assumed to be a neo-Hookean material ($\beta = -1$).}
  \label{Experimental Data Boxplot}
\end{figure}

As discussed in Section \ref{Experimental Validation Section}, the experimental setup in Figure \ref{Experimental Set Up} was used to take measurements of wave speed from a synthetic tissue sample for four levels of pre-stretch: 1.12, 1.19, 1.22 and 1.27. For each value of pre-stretch, five measurements were performed. The distribution of wave speeds for different levels of stretch can be seen in Figure \ref{Experimental Data Boxplot}.

Overall, the wave speeds measured experimentally were consistent and repeatable as demonstrated by the tight distributions for each value of pre-stretch in Figure \ref{Experimental Data Boxplot}. The observed wave speeds were between 6.57 m/s and 9.43 m/s, consistent with the overall Rayleigh wave speed values obtained from the FE simulations, see Figure \ref{Wave Speed Distribution}. We also see that as the pre-stretch of the skin increases, the speed of the Rayleigh surface wave also increases. This behaviour is consistent with both the existing analytical solution and with our GP emulator, see Figure \ref{GP Sens Analysis:d}. Furthermore, assuming that the neo-Hookean material ($\beta = -1$) model is suitable to describe the tissue sample, that the sample has uniform density ($\rho = 1116 \, \text{kg} \, \text{m}^{-3}$), and that the average Young's modulus measured by Kho et al. is correct ($E = 146 \, \text{kPa}$) \cite{kho_mechanical_2023}, we can compare the experimentally measured wave speeds to the predictions from the pre-trained GP emulator described in Section \ref{Statistical Emulation Results Section} and the analytical solution from Equation \ref{Modified Flavin Equation}. We find that there is good agreement between the experimentally measured wave speeds and the predicted wave speeds up until the highest level of pre-stretch, where the GP emulator and the analytical solution deviate from the experimental results; see Figure \ref{Experimental Data Boxplot}.

\begin{figure}[htb]
    \centering
    \subfloat[]{\label{Agreement of Experimental Data:a}\includegraphics[width=.5\linewidth]{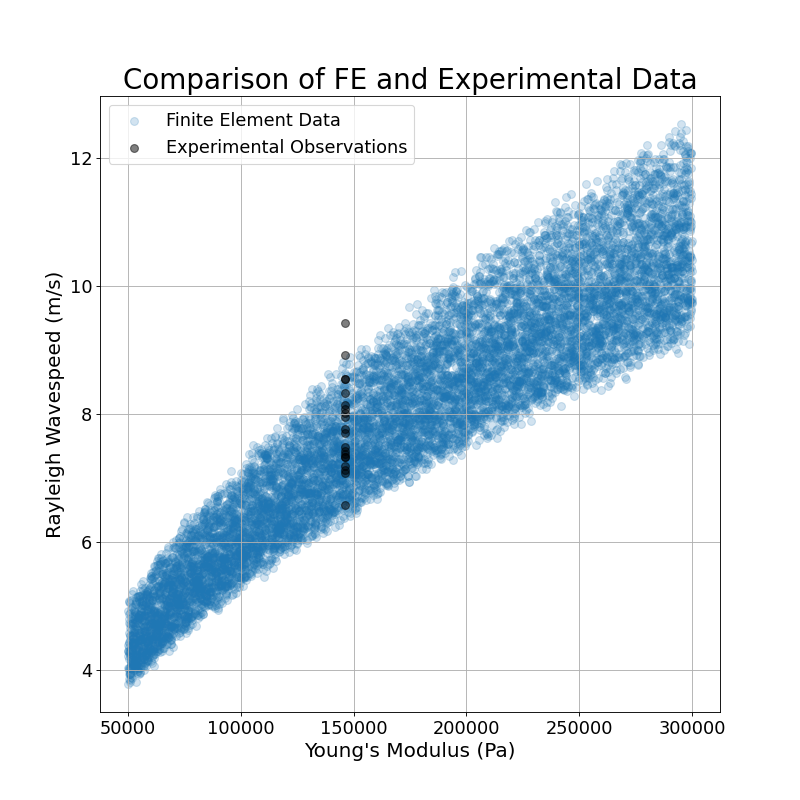}}
    \subfloat[]{\label{Agreement of Experimental Data:b}\includegraphics[width=.5\linewidth]{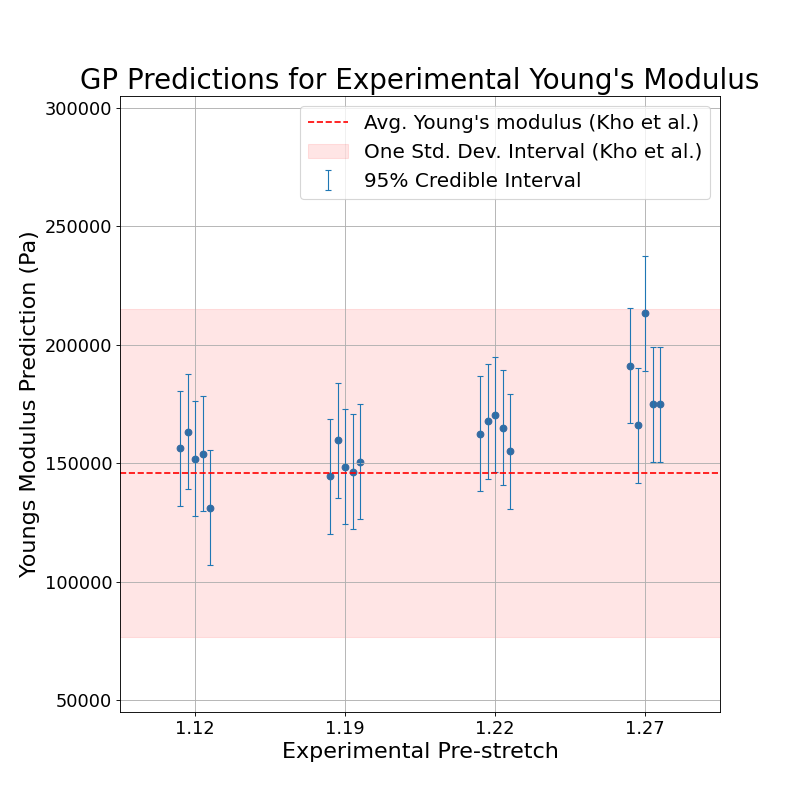}}\par\medskip
    
    \caption{(a) Scatter plot of the relationship between the Rayleigh wave speed and the Young's modulus for the FE data (blue) and the experimental observations (black). The experimental observations were all collected from synthetic tissue, with a Young's modulus of 146 kPa \cite{kho_mechanical_2023}. As shown, the experimental observations agree very well with the variation of wave speeds seen in the FE dataset. (b) Predictions from the pre-trained GP regression model using the experimental observations as inputs. Note that there is good agreement between the predicted Young's modulus from the GP model and the Young's modulus measured by Kho et al. \cite{kho_mechanical_2023}.}
    \label{Agreement of Experimental Data}
\end{figure}

We can also visualise the relationship between the Rayleigh wave speed and the Young's modulus to examine the agreement between the FE dataset and the experimental observations, see Figure \ref{Agreement of Experimental Data:a}. There is a clear positive relationship between wave speed and Young's modulus: the higher the Young's modulus, the faster the Rayleigh wave speeds. For any fixed value of Young's modulus, a $1 - 3 \, \text{m/s}$ variation in Rayleigh wave speed can be seen; this is due to different combinations of the other variables (pre-stretch, density and the material model). We see that the variation in wave speed observed in the experimental data is consistent with the variation expected from the FE data.

As discussed in Section \ref{Experimental Validation Section}, unfortunately, it was only possible to extract information about the Rayleigh wave speed, as the supersonic wave was not visible using the piezoelectric sensors employed. As such, it was not possible to validate the model presented in Section \ref{Prediction of Material Properties Section}. Instead, we use the same FE dataset to train a new GP regression model which takes in as input the pre-stretch and the Rayleigh wave speed and predicts the Young's modulus of the skin ($R^{2} = 0.9722$). Using this pre-trained model, we can input the pre-stretch and corresponding Rayleigh wave speed observations from the experimental data and obtain estimates for the Young's modulus of the synthetic skin sample, see Figure \ref{Agreement of Experimental Data:b}. The predictions from the GP regression model agree very well with the average Young's modulus of 146 kPa and fall within one standard deviation, measured by Kho et al. \cite{kho_mechanical_2023}, which we would expect to vary due to hydration levels of the sample, variations in test set up, storage conditions etc. Thus, we have demonstrated a ML model trained solely on FE simulations is capable of accurate predictions of unknown material properties of interest using  the pre-stretch and experimentally measured wave speeds from a simple wave propagation measurement.

\section{Discussion}

As demonstrated in Section \ref{Statistical Emulation Results Section}, it is possible to train an emulator to quickly and accurately produce outputs similar to a more computationally expensive simulator, using some well-chosen input values. In practice, this advance can allow for exploration of the relationship between the input and output variables (for example in the form of a sensitivity analysis), or for real-world use where relying on the simulator would be too time-consuming.

When exploring the sensitivity of the emulator to the input variables, we found that the wave speeds are by far the most sensitive to the Young's modulus $E$ and are not sensitive to the density $\rho$, see Table \ref{sens analysis regression coefficients table}. This may initially seem at odds with Equation \ref{Modified Flavin Equation}, where we see the Rayleigh wave speed is directly proportional to $\sqrt{E/\rho}$. However, recall that when sampling from the input space (see Section \ref{Input Space Sampling Section}) we allowed $E$ to have a much larger variation than $\rho$ (which in the literature is often taken to be constant), which is likely why for this emulator, the density $\rho$ is not an important predictor.

In Section \ref{Material Property Prediction Results Section}, we demonstrated that using a relatively simple GP model, it is possible to take wave speed data, which is easily accessible \textit{in vivo}, and use it to accurately predict important material properties of the skin like the stress and the pre-stretch. While this model uses only two-dimensional FE data, it could be extended, incorporating three-dimensional FE and experimental data to allow for more accurate predictions of material properties \textit{in vivo}. Given that the input parameters can be obtained easily and non-invasively, this investigation offers a method which could be of real value to surgeons and patients.

While our current dataset is based on simulated data, through this simulation approach, a number of implications for future physical measurements were identified:
\begin{enumerate}
    \item We require at least one receiver, located a known distance away from the wave generating perturbation. However, more than one receiver is preferable and allows us to not be concerned with the specifics of the wave generation process.
    \item We require enough sensitivity to be able to clearly identify the Rayleigh wave speed and ideally also the supersonic shear wave speed, which can give significant additional predictive power for $S_{11}$ and $\lambda_1$.
    \item Ideally, wave reflections should be identifiable and should not interfere with the supersonic and Rayleigh wave peak location.
\end{enumerate}

In practice, this ML method could be viewed as an alternative to inverse FE procedures which are a popular framework used in the literature to identify material properties \cite{nieuwstadt_carotid_2015} \cite{baldewsing_inverse_2008} \cite{pant_imaged-based_2017}. Regular ``forward'' FE modeling involves inputting material properties, defining boundary conditions and loading conditions, etc., and receiving some output of interest. In the inverse FE framework, some output of interest is measured/obtained and the unknown value of interest is the material parameters. These quantities are often obtained through iterative rounds of FE simulations where the material parameters are continuously tuned to minimise a predefined objective function to replicate the experimentally measured output \cite{narayanan_inverse_2021}. The benefit of our approach is that all computational complexity is ``front-loaded'': once the input space has been explored, the FE simulations performed and the ML model trained, all subsequent predictions have extremely low computational cost. This makes our method much more suitable for real-time use where, for example, a trained and validated model could be deployed in a clinic where wave speed measurements can be taken and a near instantaneous prediction of material properties can be obtained with minimal expertise required. By contrast, with the inverse FE framework, any new observation requires new rounds of potentially computationally expensive FE simulations, and significant expertise to implement and interpret them.

There are a number of limitations to the method. First and foremost is that the trained model can only make predictions in the input space it has been trained on. As such, selection of the boundaries for the input space is of paramount importance when training a model that will be deployed in the real-world. Another limitation to the models presented here is that they have been trained entirely on simulated two-dimensional FE data. For this reason, there is no noise in the training dataset and the data generation process is entirely deterministic (i.e. multiple FE simulation runs with the same set of input variables will produce identical outputs). Thus, the models presented here may not generalise well to real-world observations where the data collection process is not perfect and the physical device used may have reduced resolution. The models presented here then should serve as a proof of concept and procedural outline by which models of this kind can be employed. In practice, once some experimental \textit{in vivo} data has been collected, this simulated training dataset could be altered by reducing the precision and/or adding noise to make it more similar to the experimental observations. A more sophisticated model could then be developed by training on a combination of the adjusted FE data and experimental observations which would likely generalise much better to the experimental data. 

In this study, two different material models (neo-Hookean and Mooney-Rivlin), commonly used to model breast and other soft tissues, were used to examine if the ML model could still achieve high predictive power while trained on a mix of material models. This proved to be a success: while models trained and tested on just the neo-Hookean or just the Mooney-Rivlin subjects had slightly higher performance, the model trained on the combined dataset still had very strong predictive power. The models presented here could be extended to cover additional material models by extending the training dataset with additional simulator runs using the new material model. The Mooney-Rivlin constitutive model was selected for this study as is the simplest model that depends on both the first and second strain invariants of the left Cauchy-Green tensor ($\bar{I}_1$ and $\bar{I}_2$). Thereby, it can often provide a good fit to experimental data whilst still retaining simplicity and requiring the fitting of fewer parameters.

To demonstrate that wave speed measurements could be obtained cheaply and easily and be used to infer information about the pre-stretch, a custom  experimental rig was designed consisting of a uniaxial tensioner, a spring-loaded device to impact the surface of the skin and two piezoelectric sensors to record the waveforms. As discussed in Section \ref{Experimental Validation Results Section}, the obtained Rayleigh wave speeds were consistent, repeatable, increased as expected with increasing pre-stretch, and had good agreement with existing analytical solutions and the GP emulator. By comparing the predicted Young's modulus from a newly-trained GP regression model to the Young's modulus measured by Kho et al. \cite{kho_mechanical_2023}, see Figure \ref{Agreement of Experimental Data:b}, we demonstrate that a model trained solely on the FE simulations described in Section \ref{FE Section} is capable of fast and accurate predictions of skin material properties using wave speed observations from a simple and cheaply constructed wave propagation device.

However, it should be recognised here that there were a number of limitations to the experimental set up. Firstly, the obtained waveforms only captured the Rayleigh surface wave and the supersonic shear wave was not visible. This may be due to the sensitivity of the piezoelectric sensors and/or attenuation of the supersonic wave before arriving at the sensors which was significantly further away from the impact location than in the FE simulation. It is also possible that the spring loaded device used to generate the wave acted to increase attenuation of the perturbation. As such, it was not possible to validate the GP model presented in Section \ref{Material Property Prediction Results Section} and instead a new model was trained to demonstrate feasability. This model used the pre-stretch and Rayleigh wave speed as inputs to predict the Young's modulus $E$. It should also be noted that while the Young's modulus prediction and general agreement between the experimental data and the GP emulator were very good for low/medium pre-stretch configurations, the wave speeds obtained from the higher stretch values, especially 1.27, were higher than expected. There are a number of experimental effects which may have led to this deviation including de-lamination of the sensors from the skin surface or heterogeneous areas of strain at high stretches. Future work could involve a more thorough exploration of experimental wave speeds for different synthetic skin samples with different stiffnesses to quantify the effect that each material parameter has on the measured wave speed.

Finally, as discussed above, to generate the training data used in this study, a simplified two-dimensional FE simulation of fully elastic skin was used. Future work should involve extending this method into three dimensions.

\section{Conclusions}

In this study, a novel procedure was developed that provides real time non-invasive access to \textit{in vivo} stretch and stress. First, a simplified FE model was developed to simulate surface wave propagation in \textit{in vivo} skin. Then, a large dataset consisting of simulated real-world wave propagation experiments was constructed using the FE simulator. Using this dataset, a Gaussian process regression model was trained as an emulator that can replicate the FE model outputs with an average $R^2 = 0.9993$ at a 4 order of magnitude reduction in computational complexity. This allowed for sensitivity analysis of the physical parameters that affect the Rayleigh wave speed and supersonic shear wave speed. Then, a Gaussian process regression model was trained to solve the inverse problem of predicting clinically important material properties like the stress and the pre-stretch of \textit{in vivo} skin using measurements of Rayleigh and supersonic wave speeds. This model was found to have an average $R^2 = 0.9570$ and furthermore it was possible to interpolate simplified parametric equations to calculate the stress and the pre-stretch. Finally, an experimental device consisting of two piezoelectric sensors and a spring-loaded impactor was used to take wave speed measurements at different degrees of pre-stretch from a sample of synthetic skin tissue (Simulab). These experimental wave speeds were shown to agree well with the existing analytical solution and the Gaussian process emulator. Furthermore, a ML model trained on just the FE data was capable of taking the experimental wave speeds as inputs and predicting a Young's modulus similar to that obtained from destructive mechanical characterisation tests. These results indicate that measuring surface wave speeds to predict skin pre-stretch and stress is a feasible method which could be employed in clinical settings to inform surgical procedures.





\section*{Acknowledgments}
This publication has emanated from research supported in part by a Grant from Science Foundation Ireland under Grant number 18/CRT/6049. The opinions, findings and conclusions or recommendations expressed in this material are those of the authors and do not necessarily reflect the views of the Science Foundation Ireland.

This work is partially supported by a Government of Ireland Postdoctoral Fellowship from the Irish Research Council (Project ID: GOIPD/2022/367).

The authors wish to acknowledge the Irish Centre for High-End Computing (ICHEC) for the provision of computational facilities and support.

Part of the graphical abstract was created with BioRender.com.

\section*{Conflict of Interest}

The authors declare that there is no conflict of
interest.

\bibliographystyle{elsarticle-num} 
\bibliography{ZoteroBibTeX}






\end{document}